\documentclass[aps,prf,onecolumn,a4paper,showpacs,longbibliography]{revtex4-2}

\usepackage{bm}
\usepackage{siunitx}

\usepackage{graphicx}%
\usepackage{multirow}%
\usepackage{amsmath,amssymb,amsfonts}%
\usepackage{amsthm}%
\usepackage{mathrsfs}%
\usepackage[title]{appendix}%
\usepackage{xcolor}%
\usepackage{hyperref}
\usepackage{soul}

\graphicspath{{FIG_OLD}{FIG}}
\begin{document}
\title{Collective alignment controls rotation frustration in granular flows of elongated particles}

\author{Antonio Pol$^1$, Riccardo Artoni$^2$, Patrick Richard$^2$,} 
\affiliation{
$^1$ IATE, Univ. Montpellier, INRAE, Institut Agro, 34060,Montpellier, France\\
$^2$ MAST-GPEM, Universit\'e Gustave Eiffel, 44344 Bouguenais, France}

\date{\today}
\begin{abstract}
Dense granular flows made of elongated particles exhibit a strong inhibition of particle rotation compared to spherical grains, but the mechanisms responsible for this effect remain unclear. Using three-dimensional discrete element simulations, we investigate the angular dynamics of elongated particles in dense, confined shear flows. We systematically vary particle aspect ratio, interparticle friction, and boundary conditions to elucidate their respective roles.
We show that the reduction of the average angular velocity cannot be attributed to particle shape, friction, or solid fraction alone. Instead, it  is  controlled by the degree of collective alignment developed under shear, quantified by a nematic order parameter. Based on this observation, we propose a simple scaling law linking the average angular velocity to the local shear rate through a hampering parameter that depends solely on the orientational order via the nematic order parameter. This scaling successfully collapses data obtained for different particle properties (shape, friction), different flow patterns, and, remarkably, remains valid for two additional flow configurations.
\end{abstract}
\maketitle

\section{Introduction}\label{sec1}

Granular systems of non-spherical particles are encountered in a wide variety of natural and industrial situations, from geophysical flows to agricultural, pharmaceutical, and chemical processes. In such systems, particle shape is known to play a major role in determining both the macroscopic properties and the underlying microstructure~\cite{campbell2011,borzsonyi2012_prl, borzsonyi2013_reviewSoftMatter, Hidalgo_2018,Nagy2020, pol2022_NJOP,Fan_2024}. In particular, elongated grains exhibit behaviors that markedly differ from those of spherical particles, including particle alignment~\cite{Ribiere_2005,borzsonyi2012_prl,borzsonyi2012_pre,Borzsonyi_2016,Berzi_2016,Guillard2017,Nagy2017,Hidalgo_2018,rahim2024,Berzi2025,Amereh2026}, anisotropic force networks~\cite{Hidalgo_2009,azema2010,Azema2012}, and complex rotational dynamics~\cite{borzsonyi2012_prl,borzsonyi2013_reviewSoftMatter,artoni2019_micropolar,pol2022_NJOP,pol2023_prf}.\\
When subjected to shear, elongated particles tend to align along a preferential direction, a phenomenon that has been widely reported in experiments and numerical simulations~\cite{borzsonyi2012_prl, borzsonyi2013_reviewSoftMatter, Hidalgo_2018, pol2022_NJOP}. This alignment is accompanied by a strong correlation between particle orientation and angular velocity \cite{borzsonyi2012_pre,artoni2019_micropolar,pol2022_NJOP}, reminiscent of the Jeffery's orbits describing the motion of isolated anisotropic particles in viscous shear flows~\cite{jeffery1922}. Dense granular flows of anisotropic particles, however, differ fundamentally from such dilute suspensions. Enduring contacts, excluded-volume constraints, and frictional interactions introduce collective effects that strongly modify both the orientational statistics and the rotational dynamics of the grains~\cite{borzsonyi2012_prl,borzsonyi2012_pre,artoni2019_micropolar, pol2022_NJOP}.\\
One striking manifestation of these collective effects is the strong inhibition of particle rotation observed in dense flows of elongated grains. Unlike spherical particles, whose average angular velocity is set by the local vorticity of the flow, elongated particles rotate significantly more slowly, with an inhibition that increases with particle elongation~\cite{mandal2016,artoni2019_micropolar,pol2022_NJOP,pol2023_prf,liu2024}. While this behavior is often qualitatively attributed to steric hindrance or alignment effects, a quantitative description of rotation frustration that remains valid across particle properties and flow conditions is still missing. In particular, it is not clear whether rotation inhibition is primarily controlled by particle shape, packing density, frictional properties, or by the collective orientational order that develops under shear.\\
Another important issue concerns the role of flow geometry and boundary conditions. Confined granular flows are well known to exhibit strong shear localization and heterogeneous deformation~\cite{Taberlet2003,jop2005,Taberlet2008,Richard2008,artoni2015_prl, richard2020_gm}. It is therefore natural to ask whether the orientational state and rotational dynamics of elongated particles depend on the macroscopic organization of the flow, or whether they are instead intrinsic properties of the particles and their interactions once a steady state is reached.\\
In this work, we address these questions by means of discrete element method (DEM) simulations of dense, confined granular flows composed of elongated particles. We investigate systematically the coupling between particle rotation, local shear rate, and orientational order over a broad range of particle aspect ratios, friction coefficients, and boundary conditions. We show that rotation frustration cannot be accounted for by particle shape, friction, or solid fraction alone, but is instead governed by the degree of collective alignment in the material, quantified by a nematic order parameter. Building on this observation, we introduce a simple scaling law that relates the average angular velocity to the local shear rate through a hampering parameter depending solely on the orientational order. Remarkably, this scaling collapses data obtained for different particle properties (aspect ratio, interparticle friction), confinement conditions, and flow configurations. Moreover, we find that both the steady state orientational distribution and the correlation between particle orientation and angular velocity are essentially independent of the flow geometry, indicating that shear-induced alignment is primarily a particle property rather than a consequence of the macroscopic flow pattern.\\
The paper is organized as follows. Section~\ref{sec2:method} describes the numerical method and the flow configuration. In Sec.~\ref{sec3:res}, we present the results, focusing first on the influence of particle aspect ratio and then on the role of interparticle friction and boundary conditions. Section~\ref{sec:discussion} discusses the physical implications of these findings, their generality across different flow configurations, and their relevance for continuum descriptions of granular flows. Finally, Sec.~\ref{sec:conclu} summarizes the main conclusions and perspectives.

\section{Methodology}\label{sec2:method}

\begin{figure}
\centering
\includegraphics[width=0.45\columnwidth]{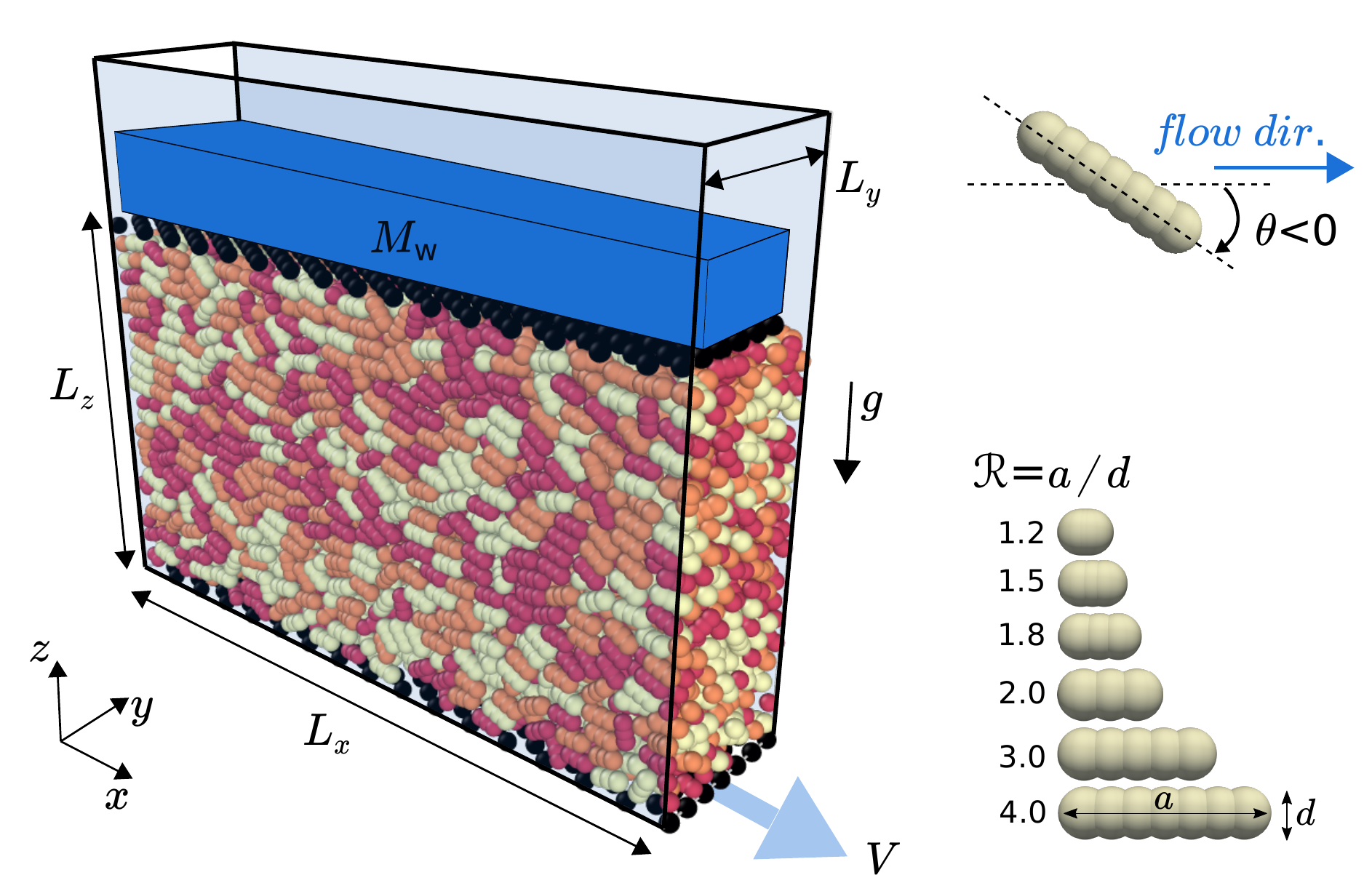}
\caption{\label{fig:fig1}
Typical geometry of the DEM simulations and of the particle shapes characterized by their aspect ratio $\mathcal{R}=a/d$. The flow configuration is a shear cell made of bumpy bottom and top walls and two sidewalls separated by a gap $L_y$. The shear is ensured by the bumpy bottom which moves at a constant velocity $V$ along the $x-$direction. The top wall is a bumpy heavy wall of mass $M_w$.}
\end{figure}
We perform 3d discrete element simulations of dense and fully confined granular flows with the open-source molecular dynamics code LIGGGHTS~\cite{liggghts2012}. The granular medium is composed of $N_p$ elongated particles with diameter $d$, length $a$ and aspect ratio $\mathcal{R}=a/d$. Different aspect ratios are considered ranging from $\mathcal{R}=1.2$ to $\mathcal{R}=4$. The particles are clumps of spheres with a minimum overlap of $d/2$ (overlap is larger for $\mathcal{R}<2$) and a polydispersity of $\pm 0.05d$ (see Fig. \ref{fig:fig1}). Preliminary analyses have shown that using larger overlaps between the clump's elements gives similar results.
Particle density is set equal to $\rho=6/\pi$ in order to have a unitary mass $m$ for spherical particles. The mass of an elongated particle is $m_p$ and the total mass of the granular medium is $M_g=N_p m_p$.
The interaction between two particles is ruled by a linear spring-dashpot model in the normal direction ($F_n=k_n\delta_n-\gamma_n\dot{\delta}_n$), $k_n$ and $\gamma_n$ being respectively the stiffness of the spring and the viscosity of the dashpot, $\delta_n$ the overlap and $\dot{\delta}_n$ the normal component of the relative velocity of the particles. The stiffness $k_n$ is taken equal to $5\times 10^6~mg/d$ and $\gamma_n$ is set in order to have a normal restitution coefficient of 0.7. 
Tangential force is ruled by a linear spring model and is limited by a Coulomb plastic condition with friction coefficient $\mu_{p}$ ($F_t=\min{(k_t \delta_t,\mu_{p}F_n)}$), where $k_t=2k_n/7$ is the tangential contact stiffness and $\delta_t$ the elastic tangential displacement between the particles.

\begin{table}[!ht]
    \centering
    \begin{tabular}{c | c | c}
        \hline
        Aspect Ratio $\mathcal{R}$ & Number of Particles $N_p$ & Mass Ratio $m_p/m$\\
        \hline
        1.2  &  4620 & 1.30 \\
        1.5  & 4300  & 1.74\\
        1.8  & 4210  & 2.14\\
        2.0  & 4200  & 2.38\\
        3.0  & 4000  & 3.75\\ 
        4.0  &  3900 & 5.13 \\
        \hline
    \end{tabular}
   \caption{Number of particles $N_p$ and mass ratio $m_p/m$ (where $m_p$ is the mass of an elongated particle and $m$ the mass of a spherical one) used in the simulations.
   For the case $\mathcal{R}=2$ the particle number refers to the base case with a cell width $L_y=10d$.}
    \label{tab:data}
\end{table}

The flow configuration (Fig. \ref{fig:fig1}) is a rectangular cuboid ($L_x=20a$, $L_y=10d$ and variable height $L_z$) and the granular medium is confined by two flat but frictional walls laterally (sidewalls), and by a top and a bottom bumpy walls composed by a regular pattern of spherical particles of diameter $d$ (triangular mesh with a spacing of $1.5d$) vertically. The contact between a particle and the sidewalls is treated in the same manner as a particle-particle contact, but with a different friction coefficient ($\mu_{w}$).
Periodic boundary conditions are applied along the flow direction, \textit{i.e.} $x$ direction.
The total number of particles $N_p$ composing the flow is adjusted in order to have a fixed ratio $M_g/L_x L_y$, independently of the particle elongation (see Table~\ref{tab:data}).

The flow is driven by the motion of the bottom wall at velocity $V$ and the system is submitted to gravity $g$. In our flow configuration, $x$ is the flow (streamwise) direction, $y$ the direction normal to the sidewalls, $z$ the (vertical) direction along which the gravity acts.
The top wall is fixed along the $x$ and $y$ directions but can adjust its $z$ position according to the balance between its weight ($M_w g$) and the force exerted by the flow. This allows us to control the vertical pressure imposed to the flow by simply changing the mass $M_w$ of the top wall. The bottom wall velocity and top wall mass are made dimensionless by considering $\tilde{V}=V/\sqrt{gd}$ and $\tilde{M}=M_w/M_g$, respectively. The adoption of this peculiar flow configuration is driven by two main reasons. First, it allows us to obtain highly heterogeneous flow conditions, where the shear rate may vary over several orders of magnitude within the same flow. This is particularly interesting for studying the rotational behavior of elongated particles, which is known to be inhibited compared to that of spherical
particles, but still linked to local shear rate.
Second, this flow configuration enables control over the flow pattern, with the potential emergence of shear localization and creep-like zones, through modification of the boundary conditions (particle–wall friction coefficient, cell width, top pressure). This allows us to study the potential influence of the flow pattern on the orientational behavior of elongated particles.

When flowing, elongated particles progressively develop a preferred orientation \cite{borzsonyi2012_prl,borzsonyi2012_pre,borzsonyi2013_reviewSoftMatter,pol2022_NJOP}. In the simulations we perform a first phase in which we let the medium to flow for a minimum local deformation of $\gamma=\dot{\gamma}(z)\Delta t>5$, where $\dot{\gamma}(z)=\partial{v}/\partial{z}$ is the local shear rate (shear rate is heterogeneous in our flow configuration), $v$ is the translational velocity along the $x$ direction, and $\Delta t$ the flow time before computing kinematic quantities. Note that for orientational related quantities we use a more strict criterion by imposing $\gamma>10$. This ensures that the medium has lost memory of the initial state and that a steady state in orientational terms has been reached (see Appendix~\ref{app:appA}). In this work, we present steady state average quantities computed by performing time averages on slices of thickness $2d$ in the $z$ direction and considering only the particles that are in the bulk of the system (vertical region characterized by $1.5 d \leq y \le L_y - 1.5 d$) to avoid effects due to the possible ordering of the particles near the sidewalls.
In this work, we vary the particle aspect ratio $\mathcal{R}$, the particle-particle and particle-wall friction coefficient $\mu_p$ and $\mu_w$ respectively, the cell width $L_y$ and the vertical confinement $\tilde{M}$, with respect to a base case configuration characterized by parameters: $\mathcal{R}=2$, $\mu_{p}=0.3$, $\mu_{w}=0.3$, $L_y=10d$, $\tilde{M}=1.2$, $\tilde{V}=1$ ($\tilde{V}=V/\sqrt{gd}$). 

\section{Results}\label{sec3:res}
\subsection{Effect of the particle's aspect ratio }\label{sec:ar_effect}

\begin{figure}
\centering
\includegraphics[width=0.48\columnwidth]{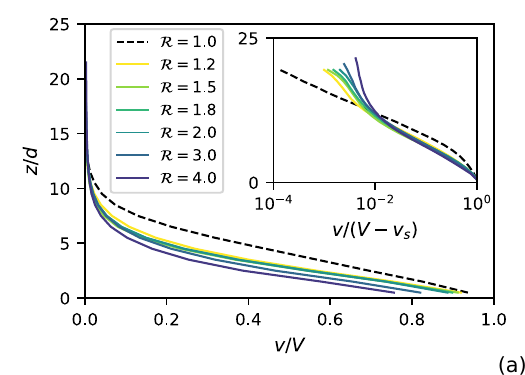}
\includegraphics[width=0.48\columnwidth]{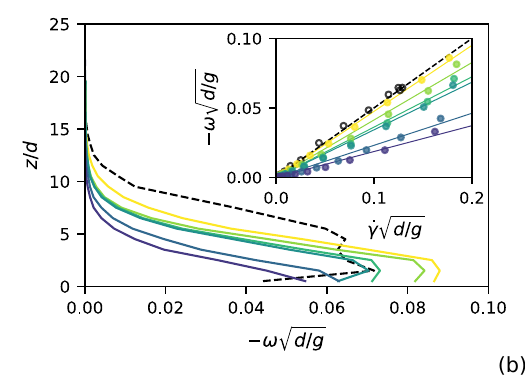}
\caption{\label{fig:fig_vel_AR}
(a) Profiles of the translational velocity for different aspect ratio $\mathcal{R}$. Inset: profiles of the rescaled velocity accounting for the slip velocity $v_s$ at the bottom wall in semilog scale. (b) Profiles of the angular velocity for different aspect ratio $\mathcal{R}$. Inset: angular velocity as a function of the local shear rate $\dot{\gamma}$. Solid lines represent the best fit of the equation $\omega = -k \dot{\gamma}$ and the dashed black line indicates the case $\omega = - \dot{\gamma}/2$ (spherical particles). The legend colors apply to both figures.
}
\end{figure}
In Fig.~\ref{fig:fig_vel_AR} we report the  profiles of the translational $v$ and angular velocity $\omega$ for different particle's aspect ratios ($1.2\leq \mathcal{R} \leq 4$). Note that $\omega$ denotes the angular velocity about the $y$ axis. We also report the case of spherical particles ($\mathcal{R}=1$) for comparison. For all the cases, we observe a clear shear localization in a band of thickness $\sim10d$ close to the bottom wall, consistently to previous works on spherical particles \cite{artoni2015_prl,richard2020_gm}. As particle elongation increases, a higher slip velocity at the wall is observed, which can be reasonably attributed to the smaller relative size of the wall asperities compared to the major axis of the particles. Rescaling the velocity profile by removing the basal slip contribution (see inset of  Fig.~\ref{fig:fig_vel_AR}a) shows that the particle elongation has only a minor effect on the velocity profile for elongated particles. Indeed, we observe only a slight change of the characteristic length of the exponential in the creep region in the upper region of the flow, while in the shear zone the velocity profiles almost collapse on a single curve, consistently with experimental findings \cite{pol2022_NJOP}.
Instead, the difference with the case of spherical particles is more evident and may be attributed to the fact that for elongated particles rotation is hindered leading to a higher mobilization of friction at the sidewalls \cite{pol2023_prf}. Additionally, this means that the shear rate profile is roughly the same for all $\mathcal{R}$ values here considered.

The influence of particle shape is instead more evident in the angular velocity profiles: longer particles rotate more slowly (Fig.~\ref{fig:fig_vel_AR}b). Remembering that the shear rate is roughly the same for all the cases with $\mathcal{R}>1$, in Fig.~\ref{fig:fig_vel_AR}b we observe a larger deviation from the classical scaling relation $ \omega  = - \dot{\gamma}/2$ (spherical particles \cite{daCruz2005,koval2009,lun1991}), as the particle elongation increases. 
Therefore, there is a clear frustration of angular motion for elongated particles. This frustration clearly depends on the particle shape (elongation), in analogy to the case of a single particle in a shear flow \cite{jeffery1922,talbot2024_pre}, but is also likely to depend on the interaction of the particle with the surrounding ones \cite{pol2022_NJOP}. This aspect will be addressed later in the paper.

\begin{figure}
\centering
\includegraphics[width=0.48\columnwidth]{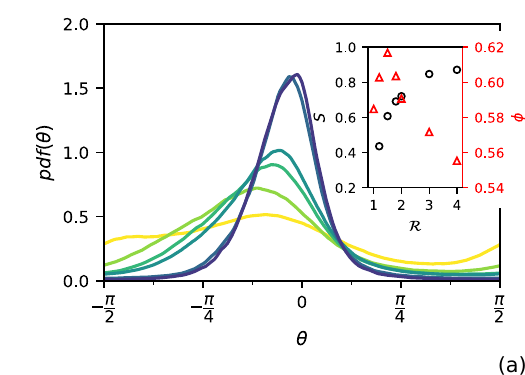}
\includegraphics[width=0.48\columnwidth]{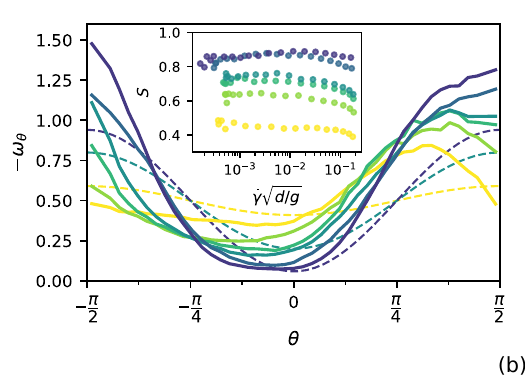}
\caption{\label{fig:oriAR}
(a) Distribution of the orientation $\theta$ of the particles for different aspect ratio $\mathcal{R}$. Inset: Order parameter $S$ ($\circ$) and average solid fraction $\phi$ ($\triangle$) as a function of the particle aspect ratio $\mathcal{R}$. (b) Normalized angular velocity of the particles as a function of the orientation $\theta$ for different aspect ratio $\mathcal{R}$.  The prediction from Jeffery's theory for $\mathcal{R}$=1.2, 2, 4 is reported for comparison (dashed lines). Inset: Local order parameter $S$ as a function of the local shear rate. The legend colors apply to both figures. 
}
\end{figure}

It is well known that elongated particles tend to show preferred orientations when submitted to a continuous shear deformation \cite{borzsonyi2012_prl,borzsonyi2012_pre,borzsonyi2013_reviewSoftMatter,pol2022_NJOP}. It is therefore interesting to investigate how a collective alignment of particles develops in our flow geometry, in which particle re-orientation can be frustrated by the highly confined nature of the chosen configuration.
Here, we focus on the orientation of the particles in the shear plane ($xz$ plane), which is characterized by the angle $\theta$ formed by the long axis of a particle and the flow direction ($x$ direction). By convention, clockwise angles are taken as negative (see Fig.~\ref{fig:fig1}). 
In Fig.~\ref{fig:oriAR}a we present the probability density function of $\theta$ for different particle elongations. 
We observe that particle elongation strongly influences both the preferential orientation and the degree of alignment of the particles. More peaky distributions are observed when increasing the particle elongation, symptom of a higher alignment of the particles, but also a lower shift of the mode of the distribution with respect to the flow direction: longer particles tend to show a lower misalignment with the flow direction.
A convenient way to measure the degree of order in our system is to use a nematic order parameter \cite{deGennes1993}:
\begin{equation}
S=\frac{1}{2N}\sum^N_{p=1}{\left(3\cos^2{\theta_p}-1\right)} ,   
\end{equation}
where $\theta_p$ is the angle between the long axis of the $p$-$th$ particle and the average angle computed over all particles with a circular average. 
As suggested by the shape of the distributions in Fig.~\ref{fig:oriAR}a, the order parameter $S$ is strongly influenced by the particle aspect ratio: it rapidly increases when increasing the particle aspect ratio in the range of small $\mathcal{R}$ and seems to tend to a constant value for sufficiently long particles (see inset in Fig.~\ref{fig:oriAR}a). This behaviour is consistent with previous works \cite{borzsonyi2012_prl,borzsonyi2012_pre,pol2022_NJOP,Nagy2017,liu2024}.
Finally, we note a slight dependence of the order parameter on the local shear rate (see inset in Fig.~\ref{fig:oriAR}b), in particular in the higher shear rate regions, for which we observe a systematic reduction of the order parameter with the shear rate. This is consistent with experimental observations \cite{pol2022_NJOP,borzsonyi2012_pre} and may be associated with the fact that particles belonging to these regions are located in the vicinity of the moving wall, which might slightly disrupt the local ordering. Only minor differences were observed in the orientation distribution of the particles if the shear zone and the creep zone were considered separately.

In previous works \cite{borzsonyi2012_pre,artoni2019_micropolar,pol2022_NJOP} it has been shown that, similarly to Jeffery's orbits, angular velocity and instantaneous particle's orientation are strongly correlated. It is therefore interesting to compute the mean angular velocity for a given particle orientation. To do so, we perform a bin averaging on the orientation angle $\theta$ of the locally normalized angular velocity $\omega/\dot{\gamma}$. This gives us a function $\omega_{\theta}$, which we can interpret as an orientational distribution of the average angular velocity. 
Practically, the trend of $\omega_{\theta}$ indicates the proneness of a particle to rotate when oriented with a given angle with respect to the shear direction. 

In Fig.~\ref{fig:oriAR}b, we observe that $\omega_{\theta}$ dramatically varies with the particle aspect ratio, with a stronger correlation between the angular velocity and the local orientation for longer particles. We also observe a marked asymmetry about $\theta=0$, which is particularly remarkable for the shorter particles. This asymmetry, which was already observed in previous experiments \cite{pol2022_NJOP}, represents a clear deviation from Jeffery’s orbit model \cite{jeffery1922}. We also note that the correlation between angular velocity and particle orientation appears stronger in our system than in the Jeffery's model (see Fig.~\ref{fig:oriAR}b). We attribute this to the granular nature of the system, in which the angular motion of a particle depends not only on the flow vorticity ($\propto \dot{\gamma}$) but also on the interaction with neighboring particles.

\begin{figure}
\centering
\includegraphics[width=0.58\columnwidth]{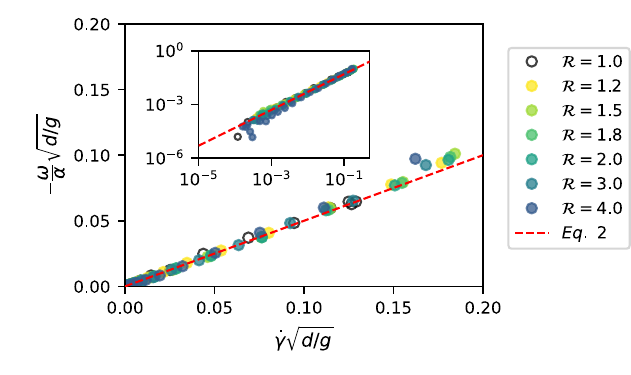}
\caption{\label{fig:om_scaled_AR}
Angular velocity rescaled with the parameter $\alpha=1-S^b$, with $b=2.6$, as a function of the local shear rate $\dot{\gamma}$. Inset: data in log-log scale.
}
\end{figure}

When discussing the kinematics of the particles, we observed a clear inhibition of the average angular motion with increasing particle aspect ratio. In parallel, we noted that the correlation between angular velocity and orientation depends on particle shape, with a dramatic increase in rotation rate for particles that are strongly misaligned with the flow direction, particularly for longer particles.
It is therefore tempting to associate this inhibition of the average angular motion with particle orientation, and more specifically with the shear alignment mechanism. On the one hand, in a medium characterized by a higher degree of order particle rotation is hampered with respect to a less ordered medium. In a highly ordered medium, in fact, particles tend to spend more time oriented with an angle $\theta$ that minimizes, on average, the angular velocity due to the $\omega$–-$\theta$ correlation. On the other hand, it is reasonable to assume that, for the same particle shape, a more ordered system is also denser and therefore the frustration to particle rotation due to the interaction with neighboring particles is stronger. In fact, unlike the case of spherical particles, the rotation of an elongated particle requires either some free space around it or the displacement/rotation of neighboring particles.
Both of these observations suggest a stronger frustration of angular motion in flows characterized by a higher order parameter $S$. In an attempt to find a relation between the angular velocity and the shear rate that applies to both spherical and elongated particles, we propose to introduce, in the classical scaling law, a hampering parameter $\alpha$ such as 

\begin{equation}\label{eq:om_gamma}
    \omega  =  -\dfrac{\alpha}{2}\dot{\gamma}.
\end{equation}

Starting from our previous assumption that angular motion is frustrated by collective alignment, we assume that the hampering parameter is a function of the order parameter $S$ and satisfies the following constraints: (i) $\alpha$ is a decreasing function of $S$; (ii) $\alpha = 1$ for a randomly oriented medium ($S=0$); (iii) $\alpha = 0$ for a perfectly ordered medium ($S=1$). Hence, we seek an expression of the form $\alpha=1-S^b$, where $b$ is a constant.
We obtain a remarkable collapse of the data on the master curve given by Eq.~\ref{eq:om_gamma} with $b=2.6$, as shown in Fig.~\ref{fig:om_scaled_AR}. This result confirms our earlier hypothesis, showing that rotations are inhibited by the collective alignment of the particles, and underlines the strong correlation between angular motion and particle orientation. We note a small deviation from the master curve for the longest particles ($\mathcal{R}=4$) for the lower shear rate values. We attribute this to size effects due to the small relative size in the $z$ direction compared to the particle length, which may be an additional source of frustration of particles rotation. We would like to emphasize that the hampering parameter $\alpha$ does not explicitly depend on the particle aspect ratio, but rather on the degree of collective alignment of the particles, quantified by the order parameter $S$. Since $S$ may depend on the shear rate, this implies that in our heterogeneous flows, characterized by large variations in shear rate, the hampering parameter may also vary along the flow height. Finally, the hampering parameter $\alpha$ unifies the case of spherical particles with the case of elongated ones. 
Indeed, for spherical particles, for which $S=0$, we trivially recover the classical relation between the average angular velocity and the shear rate. For a perfectly ordered system ($S=1$), Eq.~\ref{eq:om_gamma} predicts that the average angular velocity is zero, and therefore the particles only oscillate around the preferred orientation.

We would like to point out that this inhibition of the particle rotation cannot be easily linked to the solid fraction of the medium. Indeed, while $S$ increases monotonically with the aspect ratio of the particles, the average solid fraction $\phi$ has a non-monotonic behavior with a maximum around $\mathcal{R}=1.5$, as shown in the inset of Fig.~\ref{fig:oriAR}. This further supports the idea that the inhibition of the angular motion is strongly coupled with the collective alignment of the particles. However, for the same particle shape, it is reasonable to suppose that a higher orientational order leads to a denser system in which particle rotations are more frustrated due to the stronger entanglement with the neighboring ones, as will be discussed in Sec.~\ref{sec:sec_muP}. In this perspective, it would be interesting, in a future work, to quantify the maximum solid fraction that the medium can reach for a given particle aspect ratio and relate the hampering parameter to how far the actual solid fraction is from this maximum value.
Additionally, we have tested the relationships between shear rate and rotation rate as proposed in the recent work of Talbot et al. \cite{talbot2024_pre}, which are derived from Jeffery's orbits and a modified Jeffery model with an added noise term. 
Even though they predict an inhibition of angular motion with increasing particle elongation, these relationships do not allow us to collapse the data onto a master curve. We attribute this to the fact that they were obtained considering the case of an isolated particle in a shear flow, thus neglecting effects associated to collective particle behavior which may strengthen the inhibition of particle rotation.

The geometry of the particles has been shown to crucially influence their angular motion, in particular through the shear-alignment mechanism. At this point, it is therefore interesting to investigate how other particle properties, such as interparticle friction, and boundary conditions of the flow (interaction with the sidewalls and vertical confinement), affect the flow dynamics and particle orientation. To this end, in the remainder of the article we focus on particles with aspect ratio $\mathcal{R}=2$ and investigate the effects of the following parameters: the particle–particle friction coefficient $\mu_p$, the particle–wall friction coefficient $\mu_w$, the width of the cell $L_y$ and the vertical confinement $\tilde{M}$.


\subsection{Effect of the particle-particle friction coefficient}
\label{sec:sec_muP}
\begin{figure}
\centering
\includegraphics[width=0.48\columnwidth]{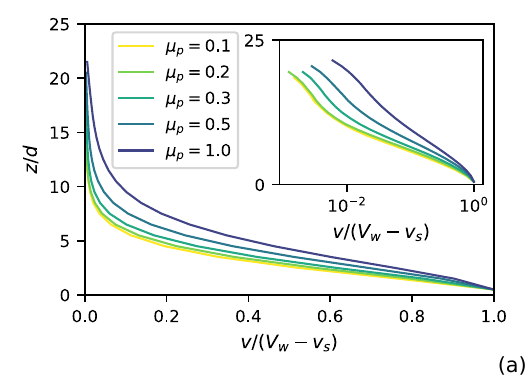}
\includegraphics[width=0.48\columnwidth]{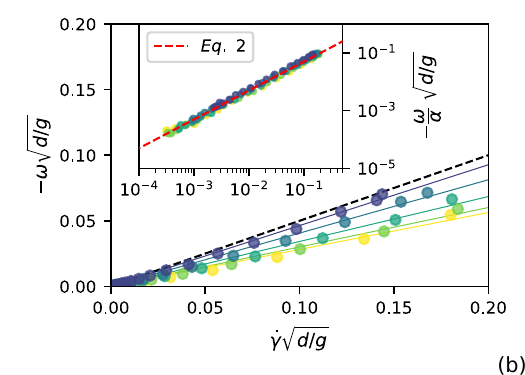}

\caption{\label{fig:fig_vel_mup}
(a) Profiles of the rescaled translational velocity for different values of the interparticle friction coefficient $\mu_p$ ($\mathcal{R}=2$). Inset: profiles of the rescaled velocity in semilog scale. (b) Angular velocity as a function of the local shear rate $\dot{\gamma}$ ($\mathcal{R}=2$). Solid lines represent the best fit of the equation $\omega = -k \dot{\gamma}$ and the dotted black line indicates the case $\omega = - \dot{\gamma}/2$ (spherical particles). Inset: angular velocity rescaled with the parameter $\alpha=1-S^b$, with $b=2.6$, as a function of the local shear rate. The colors in the legend apply to both figures.
}
\end{figure}
In order to investigate the effect of the interparticle friction we performed a set of simulations changing the friction coefficient in the range $0.1 \leq \mu_p \leq 1$. The other parameters were kept constant in all the simulations: $\mathcal{R}=2$, $\mu_w=0.3$, $\tilde{V}=1$, $\tilde{M}=1.2$. 

\begin{figure}
\centering
\includegraphics[width=0.48\columnwidth]{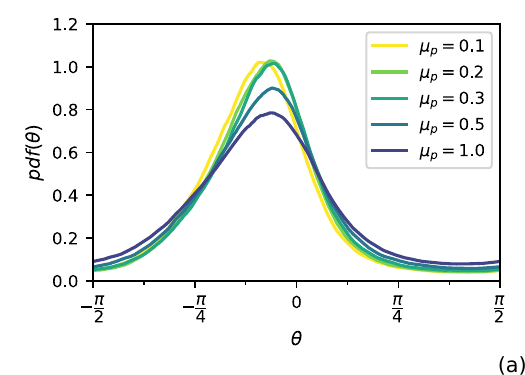}
\includegraphics[width=0.48\columnwidth]{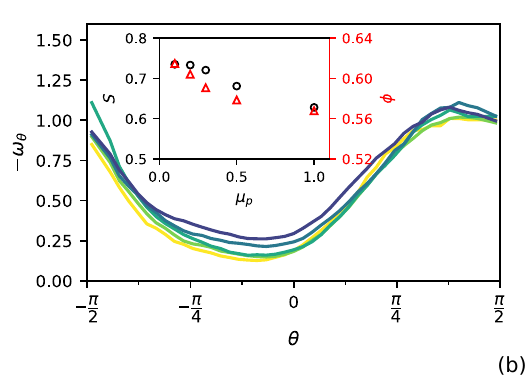}
\caption{\label{fig:ori_muP}
(a) Distribution of the orientation $\theta$ of the particles for different interparticle friction coefficient $\mu_p$ ($\mathcal{R}=2$). (b) Normalized angular velocity of the particles as a function of the orientation $\theta$ for different interparticle friction coefficient $\mu_p$ ($\mathcal{R}=2$). Inset: Order parameter $S$ ($\circ$) and average solid fraction $\phi$ ($\triangle$) as a function of the particle interparticle friction coefficient $\mu_p$. The legend applies to both figures.
}
\end{figure}
In Fig.~\ref{fig:fig_vel_mup}a we report the translational velocity profiles obtained for different values of $\mu_p$. We observe changes in the flow pattern with a milder localization of the flow and a thicker shear region for higher values of the particle-particle friction. Additionally,the shear rate reaches lower values and is more heterogeneous along the flow height for cases characterized by lower values of $\mu_p$.

We also observe that a reduction in the friction coefficient $\mu_p$ leads to a systematic inhibition of particles rotation. Indeed, Fig.~\ref{fig:fig_vel_mup}b clearly shows that, for the same shear rate, less frictional particles rotate more slowly. This is, to our knowledge, a new and interesting result, since the observed dependence of the angular motion on the frictional properties of the particles is not expected for frictional spherical particles, for which the average angular velocity $\omega=-\dot{\gamma}/2$, independently of the friction coefficient. We have performed additional simulations with spherical particles varying the particle-particle friction coefficient $\mu_p$ in the same range and we always observed a very good agreement of the data with this scaling (results not shown here). The inhibition of angular motion with interparticle friction seems therefore to be a feature of elongated particles.

According to the reasoning developed in Sec.~\ref{sec:ar_effect}, the different inhibition of angular motion observed when changing the friction coefficient implies that the correlation between $\omega$ and $\theta$, and consequently the degree of order in the medium, should also be different for the different values of the friction coefficient. This is confirmed by the distribution of particle orientation and the orientational distribution of particle angular velocity reported in Fig.~\ref{fig:ori_muP}a and b, respectively. We observe a stronger correlation between $\omega$ and $\theta$ for less frictional particles, which directly reflects on a more peaked distribution and a higher value of the order parameter $S$ (see inset in Fig.~\ref{fig:ori_muP}b). This result, allows us to test the scaling proposed in Eq.~\ref{eq:om_gamma} for the same particle geometry. From the inset of Fig.~\ref{fig:fig_vel_mup}b, it is evident that the angular velocity data collapse on a master curve when rescaled with the hampering parameter $\alpha=1-S^b$ ($b=2.6$), confirming that the frustration of the angular motion can be linked to the collective alignment of the grains. 
The reduced proneness to rotation observed for less frictional particles arises from the higher degree of order that the medium can develop. This not only results in a denser medium (see inset in Fig.~\ref{fig:ori_muP}b), where the frustration induced by neighboring particles is stronger, but also leads to a greater alignment of particles with the preferential orientation for which the angular velocity is, on average, minimized.
\subsection{Effect of the boundary conditions}\label{sec:bcs}
In the previous section, we highlighted that a change in the interparticle friction coefficient has a non negligible impact on the frustration of the angular motion and on the orientational distribution of the particles. However, $\mu_p$ was also shown to affect the kinematic profiles, so one may ask whether the impact of the interparticle friction coefficient $\mu_p$ on the $\omega-\theta$ correlation, the particles alignment and consequently on the average angular velocity, is a result of the different flow pattern, or whether it has a direct influence. To address this question, we have performed dedicated simulations in which we have modified the boundary conditions of the flow in order to obtain different flow patterns while keeping the geometry ($\mathcal{R}=2$) and the surface properties of the particles ($\mu_p=0.3$) constant.
The flow configuration adopted in this work has indeed the interesting feature of exhibiting the coexistence of several flow regimes that differ in their strain localization patterns. The origin of these regimes has been discussed in \cite{artoni2015_prl,artoni2018_jfm} and is linked to effective bulk friction heterogeneity, due to the competition between dissipation at the sidewalls and in the bulk of the flow. This competition mechanism is mainly controlled by the system’s boundary conditions, namely the particle–wall friction coefficient $\mu_w$, the confining pressure $\tilde{M}$ at the top of the cell, and the cell width $L_y$. These three quantities will be varied one at a time, and their effect on the flow will be discussed in the remainder of the paper.

\begin{figure}
\centering
\includegraphics[width=0.48\columnwidth]{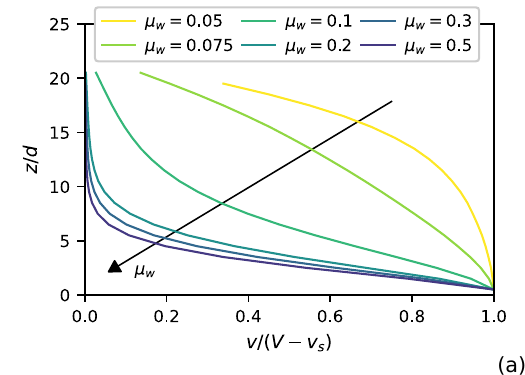}
\includegraphics[width=0.48\columnwidth]{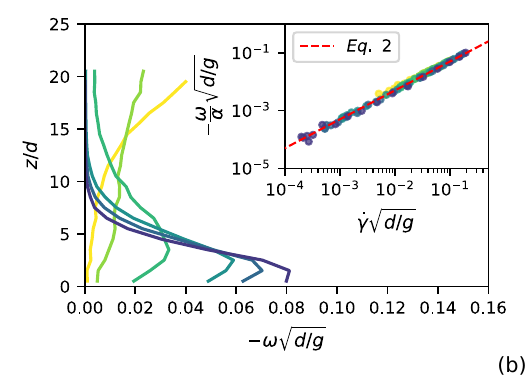}

\caption{\label{fig:fig_vel_muw}
(a) Profiles of the rescaled translational velocity for different values of the particle-wall friction coefficient $\mu_w$ ($\mathcal{R}=2$). (b) 
Profiles of the angular velocity for different values of the particle-wall friction coefficient $\mu_w$ ($\mathcal{R}=2$). Inset: Angular velocity rescaled with the parameter $\alpha=1-S^b$, with $b=2.6$, as a function of the local shear rate. The legend colors apply to both figures.
}
\end{figure}

\subsubsection{Effect of the particle-wall friction coefficient}\label{sec:sec_muW}
Previous works have shown that the adopted flow configuration is very sensitive to a change in the particle-wall friction coefficient \cite{artoni2015_prl,richard2020_gm,pol2023_prf}. It was therefore natural for us to start from the particle-wall friction coefficient $\mu_w$ to investigate if the flow regime impacts the collective alignment of the particles and consequently the inhibition of particle rotation through the hampering parameter.
\begin{figure}
\centering
\includegraphics[width=0.48\columnwidth]{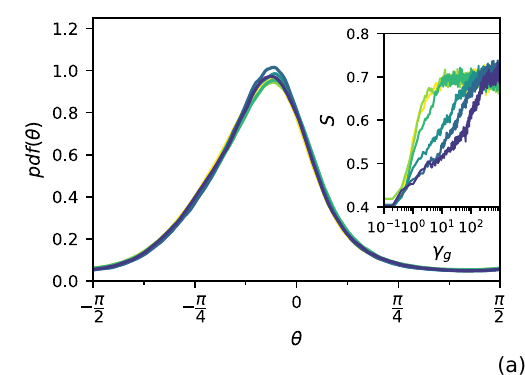}
\includegraphics[width=0.48\columnwidth]{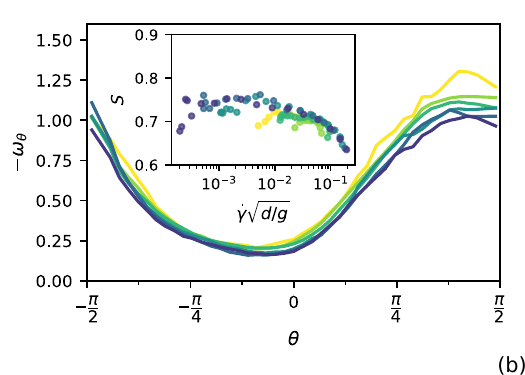}
\caption{\label{fig:ori_muW}
(a) Distribution of the orientation $\theta$ of the particles for different particle-wall friction coefficient $\mu_w$ ($\mathcal{R}=2$). Inset : Evolution of the order parameter $S$ as a function of the global strain $\gamma_g=V/L_zt$. (b) Normalized angular velocity of the particles as a function of the orientation $\theta$ for different particle-wall friction coefficient $\mu_w$ ($\mathcal{R}=2$). Inset: local order parameter $S$ as a function of the local shear rate. The colors in the legend are the same as Fig.~\ref{fig:fig_vel_muw}(a).}
\end{figure}

In Fig.~\ref{fig:fig_vel_muw}a we display the translational velocity profiles obtained for different particle-wall friction coefficients. Consistently with previous results, we can distinguish three flow regimes: (i) a shear localization in the bottom region of the flow with a creep-like flowing zone in the upper region for the higher values of $\mu_w$, (ii) a top shear localization with a bottom region which is transitioning towards a plug-like flow for the lowest value of $\mu_w$ and (iii) a nearly linear velocity profile for an intermediate value of $\mu_w$ between these two limits. It should be noted that, a fourth flow regime, characterized by a central plug with two shear zones near the bumpy walls can also be observed in this shear flow configuration \cite{artoni2015_prl}. However, we believe that we did not observe this regime because we considered only relatively thin flows (flow height of $\sim20d$). 
This modification of the flow pattern implies that angular velocity profiles are also significantly different when changing $\mu_w$ as can be noted in Fig.~\ref{fig:fig_vel_muw}b. We observe higher angular velocities in shear zones, which reflects the fact that these zones are characterized by higher shear rate values.
However, when comparing zones characterized by the same shear rate, the particles have, on average, the same angular velocity indicating that a modification in the flow pattern does not affect the proneness of a particle to rotate (not shown here). Following the scaling proposed in Eq.~\ref{eq:om_gamma}, this suggests that the hampering parameter $\alpha$, and so the order parameter $S$, are only a function of the local shear rate and independent of the flow pattern.

In this perspective, it is now interesting to focus on the particle orientation, and in particular on the possible influence of the flow regime, which we recall is strongly affected by a variation of $\mu_w$, on the shear alignment mechanism. In Fig.~\ref{fig:ori_muW}a, we observe that the distribution of particle orientation is approximately the same for all the cases regardless of the flow pattern. In relation to that, in Fig.~\ref{fig:ori_muW}b we observe that the orientational distribution of the angular velocity is also negligibly affected by a change in $\mu_w$. This independence on the flow regime, which may seem surprising at first, strongly suggests that the $\omega-\theta$ correlation and the resulting collective alignment are independent of the flow pattern and seem to be a particle property.
However, local variations of the order parameter are expected in our flows since they are characterized by a strong heterogeneity of the shear rate. In the inset of Fig.~\ref{fig:ori_muW}b we observe indeed that in regions characterized by a higher shear rate the medium is locally less ordered. This implies that the hampering parameter ($\alpha=1-S^b$) also varies along the flow height. Finally, when rescaling the angular velocity by the hampering parameter $\alpha$ we observe that the data collapse on the master curve predicted by Eq.~\ref{eq:om_gamma} as shown in the inset of Fig.~\ref{fig:fig_vel_muw}b.

We have shown that modifying the frictional interaction with the sidewalls, although significantly altering the flow pattern, does not affect the collective alignment of the particles at the steady state. In contrast, this interaction does influence the transient phase of the alignment process. We show this by reporting the evolution of the order parameter $S$ in the medium as a function of a global deformation $\gamma_g$ within the cell in the inset of Fig.~\ref{fig:ori_muW}a. This deformation is computed as $\gamma_g=\dot{\gamma}_g t$, where $\dot{\gamma}_g=V/L_z$ is a global shear rate associated to the shear cell and $t$ is the shearing time. We would like to specify that, for comparison purposes, we refer to a global shear rate $\dot{\gamma}_g$; however, the shear rate experienced by the particles depends on $\mu_w$ and may strongly vary along the flow height and between the different cases. We observe that the flow pattern, controlled here by the wall-particle friction coefficient $\mu_w$, influences the time required for the medium to reach a steady state in terms of orientational order. This behavior arises from the lower shear rate values, especially outside the shear zone, observed for higher particle-wall friction coefficients. In fact, the reaching of the steady state in orientational terms is controlled by the local deformation level, and hence requires more time in regions characterized by a lower shear rate in which particles' rotation is slower.

\subsubsection{Effect of the cell width}

\begin{figure}
\centering
\includegraphics[width=0.48\columnwidth]{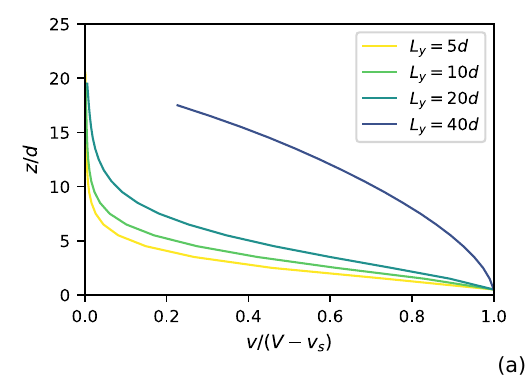}
\includegraphics[width=0.48\columnwidth]{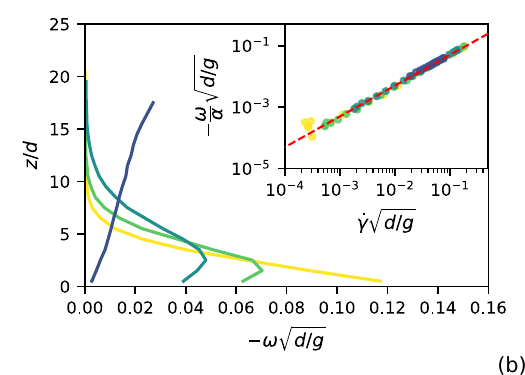}

\caption{\label{fig:fig_vel_Ly}
(a) Profiles of the rescaled translational velocity for different values of the cell width $L_y$ ($\mathcal{R}=2$). (b) 
Profiles of the angular velocity for different values of the cell width $L_y$ ($\mathcal{R}=2$). Inset: angular velocity rescaled with the parameter $\alpha=1-S^b$, with $b=2.6$, as a function of the local shear rate. The legend colors apply to both figures.
}
\end{figure}

In Sec.~\ref{sec:sec_muW}, we have discussed the crucial role on the flow regime of the frictional interaction with the sidewalls. In what follows, we present simulations in which we have varied the cell width $L_y$. We would like to underline that this is not equivalent to a reduction of the friction at the sidewalls, but rather is a complementary way to mitigate the impact of sidewalls on the flow. 

From the velocity profiles in Fig.~\ref{fig:fig_vel_Ly}a, we observe that, as one may expect, having a wider cell results in a reduction of the effect of the sidewalls. In the widest cell, we have indeed the disappearing of the bottom shear localization and a flow pattern that reminds the one observed for the lower values of the particle-wall friction coefficient. This is symptomatic of the fact that the flow is transitioning from bottom shear localization to top localization. The modification of the flow pattern reflects also on the angular velocity profiles, as can be observed in Fig.~\ref{fig:fig_vel_Ly}b. However, when comparing zones characterized by the same shear rate, the particles have, on average, the same angular velocity indicating again that a modification in the flow pattern does not affect the proneness of a particle to rotate (not shown here).

\begin{figure}
\centering
\includegraphics[width=0.48\columnwidth]{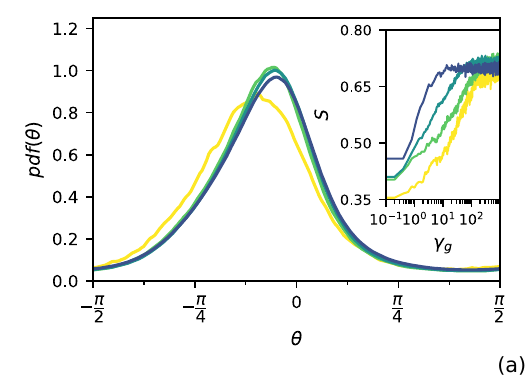}
\includegraphics[width=0.48\columnwidth]{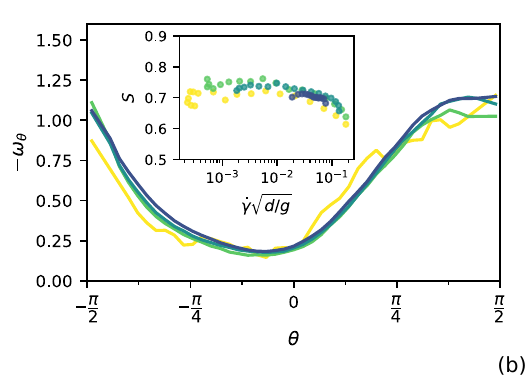}
\caption{\label{fig:ori_Ly}
(a) Distribution of the orientation $\theta$ of the particles for different cell width $L_y$ ($\mathcal{R}=2$). Inset : Evolution of the order parameter $S$ as a function of the global strain $\gamma_g=V/L_zt$. (b) Normalized angular velocity of the particles as a function of the orientation $\theta$ for different cell width $L_y$ ($\mathcal{R}=2$). Inset: local order parameter $S$ as a function of the local shear rate. The legend colors are the same as Fig.~\ref{fig:fig_vel_Ly}(a).}
\end{figure}

Consistently to what observed when changing the friction coefficient at the sidewalls, the modification of the cell width $L_y$, for a sufficiently large cell, has a negligible influence on the orientational distribution of the particles and on the $\omega-\theta$ correlation as shown in Fig.~\ref{fig:ori_Ly}a-b. This further supports the previous observation that the orientational behavior is mainly a function of the particle properties. Moreover, the angular velocity data collapse on the master curve predicted by Eq.~\ref{eq:om_gamma} when rescaled by the hampering parameter (see inset of Fig.~\ref{fig:fig_vel_Ly}b). We have observed a more marked effect on the distribution of $\theta$ and on the $\omega-\theta$ correlation for a very narrow cell ($L_y=5d$). In this case, the cell width is only a few times greater than the particle size, and the observed difference may be ascribed to finite size effects that hinder the free orientation of the particles. Additionally, we observe that, when the cell width is not excessively narrow, a modification of the boundary conditions, here the cell width $L_y$, does again influence only the transient phase of the shear alignment mechanism (see Fig.~\ref{fig:ori_Ly}a). For larger cells a faster development of the steady state is observed, which can again be related to higher values of the local shear rate along the flow height (see inset of Fig.~\ref{fig:ori_Ly}a). It should be noted that the difference in the initial value of the order parameter, is mainly related to the fact the during the gravity deposition phase, particles can more easily find a preferred orientation in a larger cell, being less frustrated by the geometric constraints imposed by the side walls.

\subsubsection{Effect of the vertical confinement}
\begin{figure}
\centering
\includegraphics[width=0.48\columnwidth]{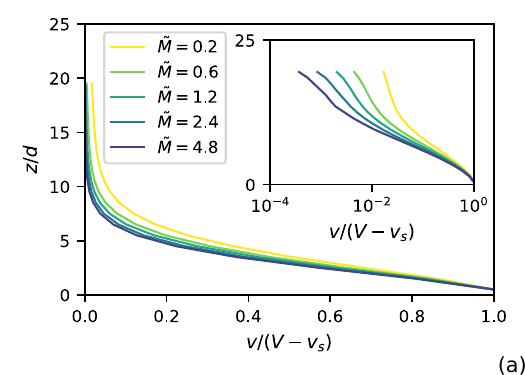}
\includegraphics[width=0.48\columnwidth]{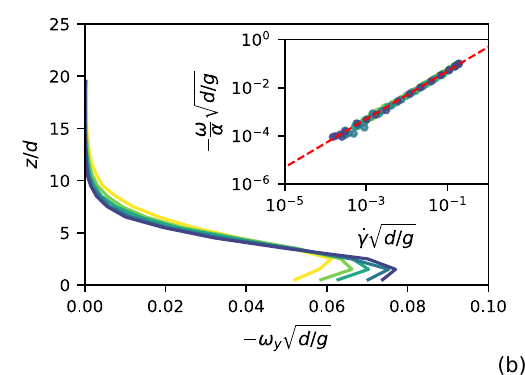}

\caption{\label{fig:fig_vel_Mtop}
(a) Profiles of the rescaled translational velocity for different values of the vertical confinement $\tilde{M}$ ($\mathcal{R}=2$).  Inset: profiles of the rescaled velocity in semilog scale. (b) Profiles of the angular velocity for different values of the vertical confinement $\tilde{M}$ ($\mathcal{R}=2$). Inset: angular velocity rescaled with the parameter $\alpha=1-S^b$, with $b=2.6$, as a function of the local shear rate. The legend colors apply to both figures.
}
\end{figure}

\begin{figure}
\centering
\includegraphics[width=0.48\columnwidth]{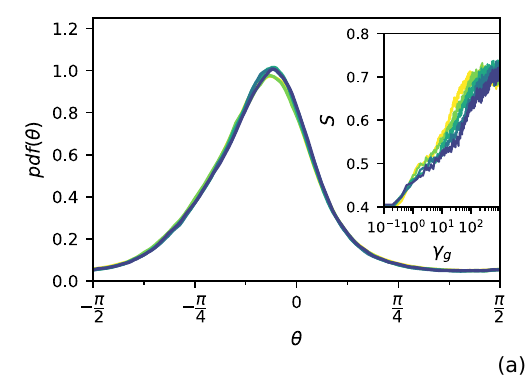}
\includegraphics[width=0.48\columnwidth]{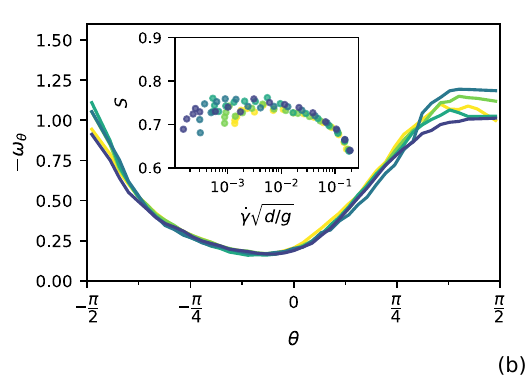}
\caption{\label{fig:ori_Mtop}
(a) Distribution of the orientation $\theta$ of the particles for different vertical confinement $\tilde{M}$ ($\mathcal{R}=2$). Inset : Evolution of the order parameter $S$ as a function of the global strain $\gamma_g=V/L_zt$. (b) Normalized angular velocity of the particles as a function of the orientation $\theta$ for different vertical confinement $\tilde{M}$ ($\mathcal{R}=2$). Inset: local order parameter $S$ as a function of the local shear rate. The legend colors are the same as Fig.~\ref{fig:fig_vel_Mtop}(a).}
\end{figure}

After having investigated the effect of the sidewalls, we now focus the influence of the vertical confinement $\tilde{M}$ imposed to the granular medium. 
The translational velocity profiles obtained for different values of $\tilde{M}$ are reported in Fig.~\ref{fig:fig_vel_Mtop}a. We observe that globally, the shear localization is less pronounced when reducing the vertical confinement, consistently with experimental observations \cite{pol2022_NJOP} and previous results obtained with spherical particles \cite{artoni2015_prl,artoni2018_jfm,richard2020_gm}. For a very low confinement, \textit{i.e.} $\tilde{M}=0.2$, we observe the formation of a region translating at an almost constant velocity above the shear region, with a non null slip velocity at the top wall. Again, we associated the fact that we do not observe the formation of a second shear region near the top wall as in \cite{artoni2015_prl} to the relatively low thickness of the flow here considered.
The angular velocity profiles are only slightly affected by the range of $\tilde{M}$ values adopted (see Fig.~\ref{fig:fig_vel_Mtop}b), with some differences notably outside the shear zone where particles rotate more slowly for a higher vertical confinement. 
At this point, it is interesting to examine the effect of $\tilde{M}$ on particle orientation. In fact, one might suppose that stronger confinement could hinder particles rotation and thus have an impact on the developing of collective alignment.
We observed that, instead, the confinement level does not affect the distribution of the particles orientation at steady state (Fig.~\ref{fig:ori_Mtop}a), but only during the transient phase of the shear-alignment process (inset of Fig.~\ref{fig:ori_Mtop}a). The transient phase lasts longer for higher vertical confinements, for which lower shear rates are observed, especially in the creep region, and particles therefore require more time to collectively align.
The $\omega-\theta$ correlation is also unaffected by the vertical confinement as shown in Fig.~\ref{fig:ori_Mtop}b, as suggested by the same orientational distribution of the particles for all the cases. Finally, in the inset in Fig.~\ref{fig:fig_vel_Mtop}b we observe that again the angular velocity values, when rescaled with the  hampering parameter $\alpha$, collapse on the master curve given by  Eq.~\ref{eq:om_gamma}.

\section{Discussion}\label{sec:discussion}

The present work provides a systematic numerical investigation of the angular kinematics and orientational order in dense, confined granular flows of elongated particles. From the detailed parametric study reported in section Sec.~\ref{sec3:res}, several general conclusions can be drawn regarding the origin of rotation frustration, the role of particle alignment, and the universality of the proposed scaling law.

One of the most robust results of this study is that, for a given particle aspect ratio $\mathcal{R}$ and interparticle friction coefficient $\mu_{p}$,  the orientational statistics of elongated particles at steady state appear to be nearly independent of the flow pattern. By varying boundary conditions such as the particle-wall friction coefficient, the cell width, or the vertical confinement, we generate markedly different flow regimes, including bottom-localized shear, top-localized shear, and nearly linear velocity profiles. 
Despite these strong differences, the distributions of particle orientation $\theta$, the $\omega$--$\theta$ correlations,
and the nematic order parameter $S$ remain almost unchanged when compared at identical local shear rates. This observation strongly suggests that shear-induced alignment is primarily a particle property, governed by particle shape and interparticle interactions, rather than by the macroscopic flow organization. The flow pattern controls where shear is localized and how fast particles move and orient, but it does not alter the intrinsic coupling between rotation and orientation once a steady state is reached.
Boundary conditions therefore affect mainly the transient dynamics of alignment, by modulating the local shear rate and hence the accumulated strain required to reach the steady state in terms of particle orientation.
\begin{figure}
\centering
\includegraphics[width=0.58\columnwidth]{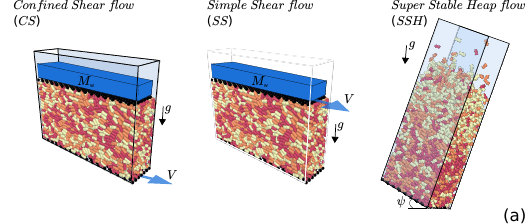}\\
\includegraphics[width=0.48\columnwidth]{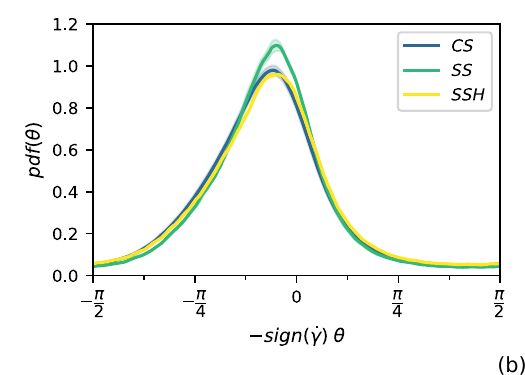}
\includegraphics[width=0.48\columnwidth]{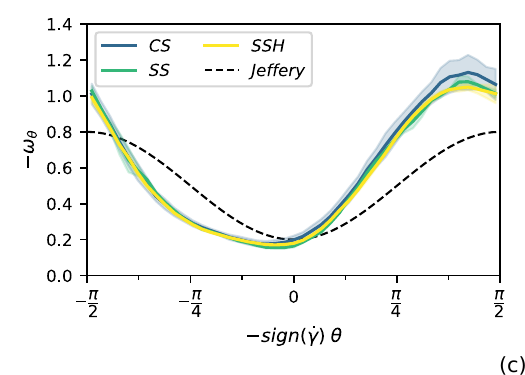}
\caption{\label{fig:comp_flows_ori}
(a) Typical geometry of the DEM simulations for the different flow configurations: Confined Shear flow (CS), Simple Shear flow (SS), Super Stable Heap flow (SSH). (b) Distribution of the orientation $\theta$ of the particles. (c) Normalized angular velocity of the particles as a function of the orientation $\theta$. The prediction from Jeffery's theory for $\mathcal{R}$=2 is reported for comparison (dashed line).
In both figure the shaded area represents the standard deviation calculated considering all cases for each flow configuration. The same particle properties are used in the three configurations ($\mathcal{R}=2$, $\mu_p=0.3$).}
\end{figure}

\begin{figure}
\centering
\includegraphics[width=0.48\columnwidth]{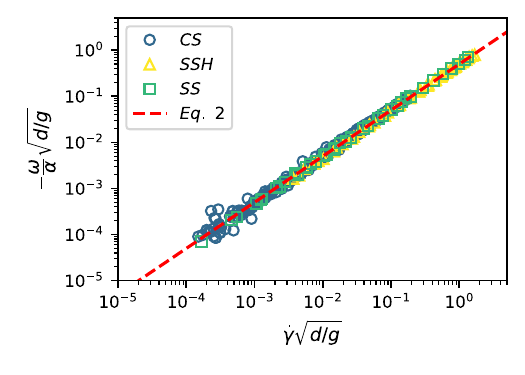}
\caption{\label{fig:comp_flows_omega}
Angular velocity rescaled with the parameter $\alpha=1-S^b$, with $b=2.6$, as a function of the local shear rate for different flow configurations: Confined Shear flow (CS), Simple Shear flow (SS), Super Stable Heap flow (SSH). The same particle properties are used in the three configurations ($\mathcal{R}=2$, $\mu_p=0.3$).
}
\end{figure}

A central result of this work is the identification of a  scaling law linking the average angular velocity $\omega$ to the local shear rate $\dot{\gamma}$ through a hampering parameter $\alpha$: $\omega = -\alpha \dot{\gamma}/2,$ with $\alpha = 1 - S^{b}$.
This relation successfully collapses data obtained for different particle aspect ratios, interparticle friction coefficients, and boundary conditions onto a single master curve.
The proposed scaling highlights that the inhibition of particle rotation is not directly controlled by particle elongation, friction, or solid fraction taken separately, but rather by the degree of collective alignment developed in the medium. In this
sense, the order parameter $S$ acts as a macroscopic descriptor of the microscopic constraints that hinder rotation. As alignment increases, particles spend more time near preferential orientations for which the instantaneous angular velocity is minimal,
and rotation events increasingly require cooperative rearrangements of neighboring particles, leading to a strong reduction of the average angular velocity.
The exponent $b \simeq 2.6$ obtained for the confined shear configuration shows a highly nonlinear coupling between orientational order and rotation frustration. While the origin of the value of $b$ remains unexplained, it likely reflects the combined effects of excluded volume, steric constraints, and the sharp peaking of the orientation distribution at large $S$. To test its generality, in addition to the confined shear flow (CS) configuration studied here, we performed simulations in two other flow configurations: a simple shear flow (SS) and a super stable heap (SSH) configuration, \textit{i.e.} a laterally confined heap flow~\cite{Taberlet2003,Taberlet2008,Richard2008,richard2020_gm} (see Fig.~\ref{fig:comp_flows_ori}a). Details about these configurations are given in Appendix~\ref{app:appA}. Interestingly, the results obtained in the SS and SSH configurations (see Fig.~\ref{fig:comp_flows_ori}) and their comparison with the results obtained above in the CS configuration show that the flow geometry has only a mild effect on the distribution of particle orientations. The distribution for the cases CS and SSH are almost identical, while we observe a slightly more peaked distribution in the SS configuration which may be symptomatic of the fact that, in this case, particles motion is not frustrated by lateral confinement. The correlation between the particle orientation and the angular velocity remains largely unaffected when changing the flow configuration. Note that, similarly to what was done for the CS flow configuration, we computed orientational related quantities with reference to the flow region for which the particles have experienced a local deformation $\gamma > 10$. Remarkably, the same functional form of the scaling law linking the average angular velocity $\omega$ to the local shear rate $\dot{\gamma}$
is recovered (see Fig.~\ref{fig:comp_flows_omega}). Preliminary simulations with a flow configuration with the same geometry as the SS one, but without gravity, have shown results that are consistent with the picture given above for both the $pdf(\theta)$ and the $\omega$-$\theta$ correlation as well as for the scaling law proposed in Eq.~\ref{eq:om_gamma}, albeit with a slightly different value of the exponent $b\approx 2.9$ (results not shown here). In the future it may be interesting to extend the analysis to other aspect ratios and particle shapes in an attempt to better understand the origin of $b$.

An important outcome of this study is that rotation frustration cannot be directly scaled with the solid fraction $\phi$ alone. The average solid fraction indeed displays a non-monotonic dependence on particle aspect ratio, whereas the order parameter increases monotonically. This decoupling shows that density is not a sufficient variable to characterize rotational hindrance across different particle shapes. We suppose that a more relevant quantity may be the distance of the actual solid fraction from the critical solid fraction at which the stress diverges in a shear flow, which is a function of particle properties such as shape and friction. This finding emphasizes that angular motion in granular flows of elongated particles is controlled by anisotropic microstructural characteristics, rather than by scalar observables such as the solid fraction $\phi$. Rotation requires either local free volume or coordinated motion of neighbors, and these constraints depend strongly on how particles are oriented relative to one another, not only on how densely they are packed. Consequently, in our opinion, any attempt to model angular kinematics using density-based rheological closures alone is bound to miss essential aspects of the physics for non-spherical grains.

The strong coupling observed between particle orientation and angular velocity has direct implications for continuum theories of granular flows. In particular, the present results naturally connect to micro-polar or Cosserat-type descriptions, in which grain
rotation and spin are treated as independent fields coupled to the velocity gradient. Our findings suggest that such theories should explicitly incorporate the correlation between rotation rate and particle orientation, possibly through an internal variable related to the nematic order parameter. In this perspective, the hampering parameter $\alpha(S)$ can be seen as an effective reduction of the local vorticity experienced by the grains due to collective alignment. Introducing such a coupling into micro-polar balance equations could provide a route toward constitutive models aiming to capture both spherical and
elongated particle flows within a unified framework.
%

\section{Conclusion}\label{sec:conclu}
In this work, we investigated the angular dynamics of dense confined granular flows composed of elongated particles subjected to shear. Using DEM simulations, we focused on the coupling between particle rotation, local shear rate, and orientational order, with the aim of identifying the mechanisms responsible for rotation frustration in non-spherical granular media.\\
Our results demonstrate that, in media composed of elongated particles, rotation is strongly inhibited and that this inhibition cannot be accounted for solely by particle shape, friction, or solid fraction. Instead, we show that rotation frustration is governed by the degree of collective alignment developed in the material, quantified by the nematic order parameter $S$. By introducing a hampering parameter that depends on $S$, we propose a simple scaling law relating the average angular velocity to the local shear rate. This scaling collapses data obtained over a wide range of aspect ratios, friction coefficients, and confinement conditions.\\
Importantly, we find that the flow geometry negligibly affects the steady-state orientational distribution and the correlation between particle orientation and angular velocity, both of which remain remarkably robust across different flow configurations. The same functional form of the scaling law is recovered in simple shear and super stable heap flows which highlights the generality of the proposed description.\\
These findings emphasize the important role of microstructural anisotropy in controlling angular kinematics in granular flows of elongated particles. They also suggest that continuum descriptions of such materials should explicitly incorporate orientational order and its coupling to grain rotation, for instance within micro-polar or Cosserat-type frameworks. More broadly, the present study provides a unified picture of rotation frustration in shape-anisotropic granular media and opens the way toward constitutive models that consistently account for particle shape effects beyond simple density-based approaches. Similar mechanisms may also be relevant in other systems involving elongated particles, such as suspensions of blood cells or nanorods.

\section*{Data availability}
The data that support the findings of this article are openly available at \href{https://doi.org/10.57745/XQB5LH}{https://doi.org/10.57745/XQB5LH}.
\section*{Acknowledgments}
This work was supported by the GLiCID High Performance Computing facility. 
Approximately 90\% of the data used in this study comes from a pre-existing dataset generated as part of a previous study \cite{pol2023_prf}. This allowed us to reduce the carbon footprint associated with the numerical simulations.
\begin{appendices}

\section{Normal to the sidewalls profiles}\label{app:appB}
In Sec.~\ref{sec:bcs}, we have shown that modifying either the particle–wall friction coefficient or the cell width strongly influences the kinematic profiles along the flow height and may induce a drastic change in the flow pattern.
The effect of the sidewalls on the flow behavior is therefore crucial, and it is of interest to analyze how this influence is reflected in profiles normal to the sidewalls (along the y direction). Additionally, variations in the particle–particle friction coefficient can influence flow localization, as shear localization is governed by the competition between dissipation at the sidewalls and within the bulk of the flow \cite{artoni2018_jfm}. For brevity, we refer to profiles normal to the sidewalls as transverse profiles. We recall that the parametric analysis was performed with respect to a base case configuration characterized by parameters: $\mathcal{R}=2$, $\mu_{p}=0.3$, $\mu_{w}=0.3$, $L_y=10d$, $\tilde{M}=1.2$, $\tilde{V}=1$.
\begin{figure}[b]
\centering
\includegraphics[width=0.68\columnwidth]{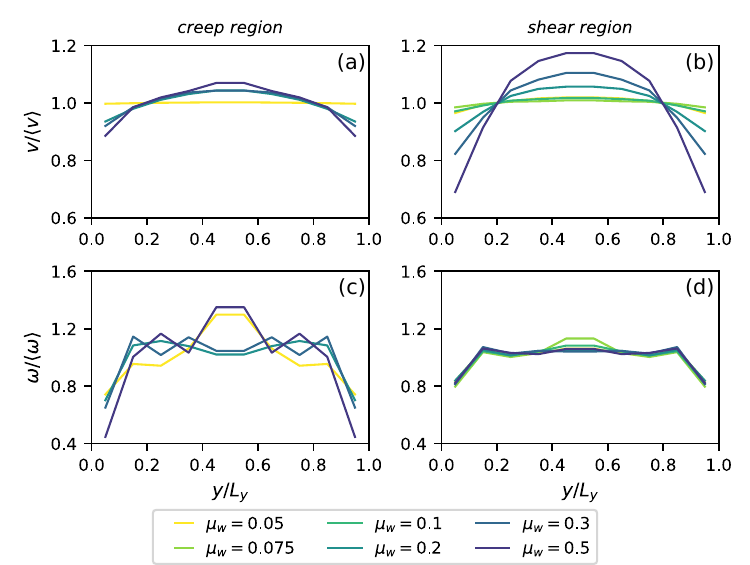}
\caption{\label{fig:appB1}
Transverse profiles of the translational velocity $v$ for different values of the particle-wall friction coefficient $\mu_w$ in (a) the creep region and (b) in the shear region (when shear localization is not observed, the entire flow height is considered as the shear region).  Transverse profiles of the angular velocity $\omega$ for different values of the particle-wall friction coefficient $\mu_w$ in (c) the creep region and (d) in the shear region. Quantities are rescaled on their averaged value in the $y$ direction. }
\end{figure}
\begin{figure}
\centering
\includegraphics[width=0.68\columnwidth]{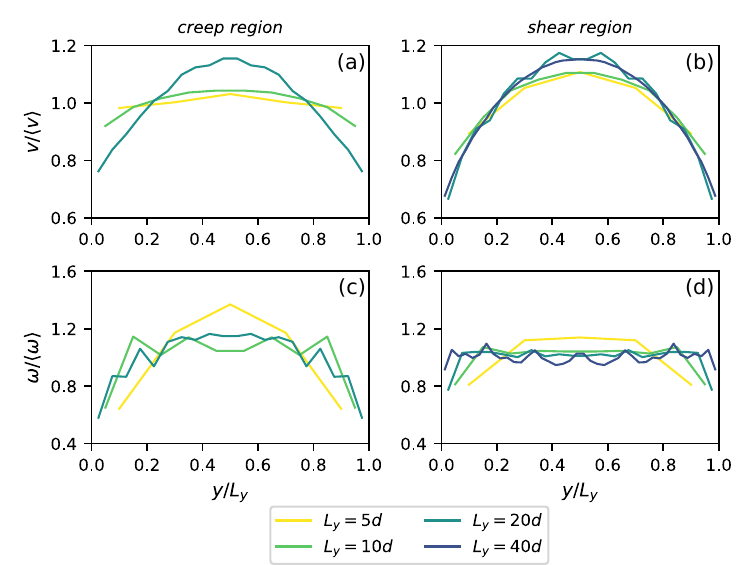}
\caption{\label{fig:appB2}
Transverse profiles of the translational velocity $v$ for different values of the cell width $L_y$ in (a) the creep region and (b) in the shear region (when shear localization is not observed, the entire flow height is considered as the shear region).  Transverse profiles of the angular velocity $\omega$ for different values of the cell width $L_y$ in (c) the creep region and (d) in the shear region. Quantities are rescaled on their averaged value in the $y$ direction. }
\end{figure}
\begin{figure}
\centering
\includegraphics[width=0.68\columnwidth]{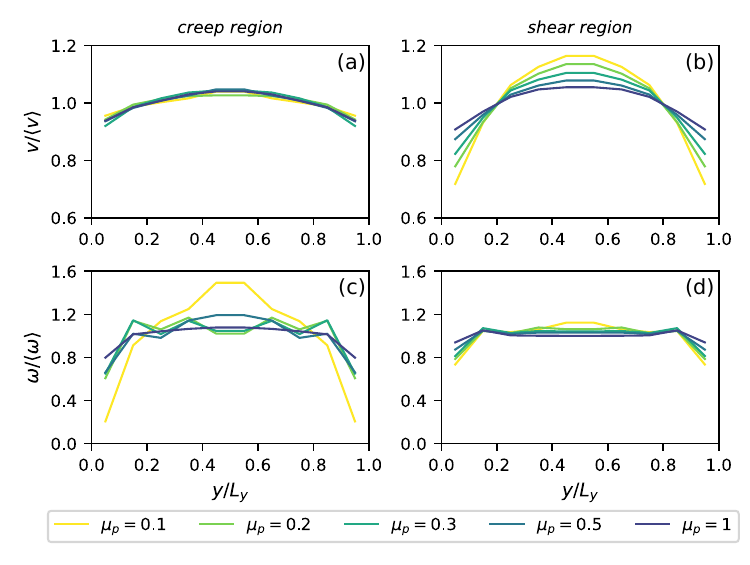}
\caption{\label{fig:appB3}
Transverse profiles of the translational velocity $v$ for different values of the particle-particle friction coefficient $\mu_p$ in (a) the creep region and (b) in the shear region (when shear localization is not observed, the entire flow height is considered as the shear region).  Transverse profiles of the angular velocity $\omega$ for different values of the particle-particle friction coefficient $\mu_p$ in (c) the creep region and (d) in the shear region. Quantities are rescaled on their averaged value in the $y$ direction.}
\end{figure}

In Fig.~\ref{fig:appB1}, we report the transverse profiles of the translational velocity $v$  (along the $x$ axis) and the angular velocity $\omega$ (about the $y$ axis) for different values of the particle–wall friction coefficient $\mu_w$. Note that the quantities are rescaled by their average value along the $y$ direction. Since the behavior of the granular medium can differ between the shear and creep regions, the quantities are computed separately for each region; these zones are identified following the criterion proposed in previous works \cite{richard2020_gm,pol2023_prf}. For cases in which no clear shear localization is observed, we consider the entire flow height as a single shear region. Focusing first on the shear region, we observe a variation of the translational velocity $v$ along the $y$ direction, indicating the presence of a transverse shear rate. This variation is more pronounced in cases where bottom shear localization is observed ($\mu_w > 0.1$) and becomes systematically stronger as the particle–wall friction coefficient increases (see Fig.~\ref{fig:appB1}b). This suggests that the granular medium feels the presence of the sidewalls over longer distances when the sidewalls are more frictional. Furthermore, the existence of a velocity gradient along the $y$ direction indicates that rotation in the $xy$ plane may not be negligible, especially in the first layers of particles near the sidewalls. Similar conclusions apply to the creep region, although the variation of $v$ is much weaker (see Fig.~\ref{fig:appB1}a). The angular velocity $\omega$, on the other hand, is approximately constant in the bulk of the flow, with only the first layers of particles near the sidewalls exhibiting a reduced angular velocity, likely due to the frustration imposed by the sidewalls (see Fig.~\ref{fig:appB1}c–d). Additionally, we note that variations in angular velocity are slightly stronger in the creep region than in the shear region.

In Fig.~\ref{fig:appB2}, we report the transverse profiles of the translational velocity $v$ and the angular velocity $\omega$ for different values of the cell width $L_y$. We observe that for relatively narrow cells ($L_y \leq 20d$), the variation of $v$ along the $y$ direction becomes stronger as $L_y$ increases. This can be attributed to the fact that the bulk of the material feels the presence of the sidewalls less in wider cells. Only minor differences are observed when increasing $L_y$ from $20d$ to $40d$; however, the flow pattern differs between these two cases, and no shear localization is observed in the larger cell (we recall that a stronger transverse velocity gradient is observed in cases with more pronounced shear localization, see Fig.~\ref{fig:appB1}b). Again, the behavior in the creep region is similar to that in the shear region, although the velocity gradient along the $y$ direction is weaker. Similarly to what is observed when varying the particle–wall friction, the angular velocity $\omega$ exhibits variations along the $y$ direction only within the first layers of particles near the sidewalls, while remaining nearly constant in the bulk.

In Fig.~\ref{fig:appB3}, we report the transverse profiles of the translational velocity $v$ and angular velocity $\omega$ for different values of the particle–particle friction coefficient $\mu_p$. We observe that, in the shear region, transverse variations of the translational velocity are systematically more pronounced for lower values of $\mu_p$, thereby inducing higher transverse shear rates at lower particle–particle friction coefficients. This may be attributed to the fact that, for lower values of $\mu_p$, effects associated with frictional interactions with the sidewalls are systematically stronger than those related to bulk friction. This is consistent with the more localized character of the flow observed for lower values of $\mu_p$ (Sec.~\ref{sec:sec_muP}). In contrast, in the creep region, the transverse velocity gradient is very mild and no significant differences are observed when varying the particle–particle friction coefficient. Overall, the angular velocity $\omega$ exhibits mild variations along the $y$ direction, primarily concentrated within the first particle layers near the sidewalls. However, for the lowest value of $\mu_p$, we observe in the creep region a strong reduction of angular velocity near the sidewalls, accompanied by an increase in the bulk.

Transverse profiles are negligibly affected by either a change of the particle aspect ratio (for $\mathcal{R}>1$) or the vertical confinement $\tilde{M}$ in the range of variation considered in this work. They are therefore not presented for these cases for a sake of brevity. Finally, we observed that the average value of the vertical shear rate ($\dot{\gamma}=\partial{v}/\partial{z}$), in both the shear and creep regions and for all the cases here studied, is approximately one order of magnitude greater than the transverse shear rate ($\dot{\gamma}_y=\partial{v}/\partial{y}$). However, velocity gradients in the transverse direction due to interactions with the sidewalls are not negligible and may induce significant phenomena, notably particle rotation in the $xy$ plane, making this an interesting direction for future work.

\section{Other flow configurations}\label{app:appA}
\begin{figure}
\centering
\includegraphics[width=0.9\columnwidth]{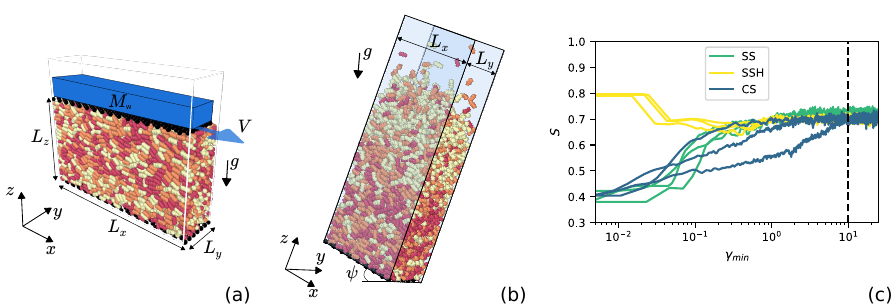}
\caption{\label{fig:appA1}
Typical geometry of the DEM simulations for the (a) Simple Shear flow configuration (SS) and for the (b) Super Stable Heap flow configuration (SSH).
(c) Evolution of the order parameter $S$ as a function of the minimum strain $\gamma_{min}=\min(\dot{\gamma})t$ for the CS, SS and SSH flow configurations. For the CS configuration we shown the cases $\mu_w=0.05,~0.2,~0.5$ which are characterized by markedly different flow patterns. The dashed line represents the threshold value of the deformation $\gamma_t=10$ adopted to define the steady state in orientational terms.}
\end{figure}

To test the generality of the scaling proposed in Eq.~\ref{eq:om_gamma}, we performed simulations in two additional flow configurations: a Simple Shear flow (SS) and a Super Stable Heap flow (SSH) as shown in Fig.~\ref{fig:appA1}a and Fig.~\ref{fig:appA1}b respectively. The SS configuration provides a flow which is not laterally bounded, while the SSH configuration allows comparison with a free-surface flow. In all these simulation, the same contact parameters as those given in Sec.~\ref{sec2:method} are used.

For the simple shear flow we use a cell of dimension: $L_x=40d$, $L_y=10d$ and variable height $L_z$, and with periodic boundary conditions along both the $x$ and $y$ directions (lateral walls are not present). The flow is bounded vertically by a top and a bottom bumpy wall composed by a regular pattern of spherical particles of diameter $d$ (triangular mesh with a spacing of $1.5d$). The flow is composed of 4200 particles of aspect ratio $\mathcal{R}=2$ with a particle-particle friction coefficient $\mu_p=0.3$.
The pressure is controlled similarly to the CS configuration by using a heavy wall whose motion is free along the $z$ direction so that its position can adjust according to the balance between its weight ($M_w g$) and the force exerted by the flow. In all the simulation we adopted a top pressure of $\tilde{M}=1.2$. In this configuration the flow is driven by the displacement of the top wall at a constant velocity $\tilde{V}$. We performed three simulation with a wall velocity of $\tilde{V}=0.1,1,10$ to cover a range of shear rate values similar to the one of the CS flow configuration.

For the super stable heap we use a cell of dimensions: $L_x = 20d$, $L_y = 10d$, while the flow is not bounded at the top. Periodic boundary conditions are imposed along the $x$ direction and the cell is inclined by an angle $\psi$ with respect to the horizontal. The flow is laterally bounded by flat but frictional sidewalls as in the CD configuration. The bumpy bottom wall is composed by a regular pattern of spherical particles of diameter $d$ (triangular mesh with a spacing of $1.5d$). The flow is composed of 6000 particles of aspect ratio $\mathcal{R}=2$ with a particle-particle friction coefficient $\mu_p=0.3$ and a particle-wall friction coefficient $\mu_w=0.3$. In this configuration, the bottom wall is kept fixed and the flow is driven by gravity. We performed three simulations with a slope angle $\psi=42^{\circ},47^{\circ},52^{\circ}$. 
The sample is prepared by pluviation of an aligned cloud of particles ($\theta=0^{\circ}$ for all the particles). Then, we perform a preliminary phase in which we let the medium flow with a high value of the slope angle ($\psi=60^{\circ}$) before reducing the slope angle to the desired value. The decision to start from an aligned assembly was made to provide further evidence that the orientation of particles in the steady state is not influenced by the initial state.
Note that, in the SSH configuration, the orientation $\theta$ of a particle is computed with respect to the flow direction, which differs from the horizontal.

For the computation of orientational related quantities we refer only to flow regions for which the particles have experienced a local deformation $\gamma(z) > \gamma_t$ with $\gamma_t=10$. This threshold on a minimal local deformation represents a robust criterion for ensuring that a steady state is achieved in orientation terms, regardless of the flow regimes and flow configurations.
To confirm the adoption of this criterion, we report in Fig.~\ref{fig:appA1}c the evolution of the order parameter $S$ with respect to the minimum deformation $\gamma_{min}$ in the medium. We compute $\gamma_{min}$ by considering the minimum shear rate value in the medium (shear rate is highly heterogeneous in our flows). We can observe that, when the medium has experienced a deformation greater than the threshold value $\gamma_t$ a steady state in orientation terms is reached independently of the flow configuration. Furthermore, this threshold is sufficient to ensure that the medium loses memory of the initial state, as indicated by an approximately equal value of the order parameter $S$, for the three flow configurations, in the steady state. Note that, for the CS and SS configurations, the medium is initially in a state of lower order than the steady state, while for the SSH configuration, it starts from a state of higher order.



\end{appendices}


\bibliography{Bib_Pol_et_al_PRF_2026}

\end{document}